\documentclass[runningheads]{llncs}
\usepackage{graphicx}
\usepackage{subcaption}
\usepackage[colorlinks,urlcolor=blue]{hyperref}
\usepackage{amssymb}
\usepackage{amsmath}

\makeatletter
\newcommand{\printfnsymbol}[1]{%
    \textsuperscript{\@fnsymbol{#1}}%
}
\makeatother

\begin{document}
\title{
    Nodule2vec: a 3D Deep Learning System for Pulmonary Nodule Retrieval
    Using Semantic Representation%
}
\titlerunning{Nodule2vec}
\author{
    Ilia Kravets\inst{1,2,}\thanks{I. Kravets, T. Heletz -- equal contribution}
    \and
    Tal Heletz\inst{1,2,}\printfnsymbol{1}
    \and
    Hayit Greenspan\inst{3}
}
\authorrunning{I. Kravets et al.}
\institute{
    Independent Researcher
    \email{\{ilia.kravets,talheletz123\}@gmail.com}
    \and
    Y-Data, Yandex School of Data Analysis, Tel Aviv, Israel
    \and
    Tel Aviv University, Tel Aviv, Israel
    \email{hayit@eng.tau.ac.il}
}
\maketitle
\begin{abstract}

Content-based retrieval supports a radiologist decision making process by presenting the
doctor the most similar cases from the database containing both historical diagnosis and
further disease development history. We present a deep learning system that transforms a
3D image of a pulmonary nodule from a CT scan into a low-dimensional embedding vector. We
demonstrate that such a vector representation preserves semantic information about the
nodule and offers a viable approach for content-based image retrieval (CBIR). We discuss
the theoretical limitations of the available datasets and overcome them by applying
transfer learning of the state-of-the-art lung nodule detection model. We evaluate the
system using the LIDC-IDRI dataset of thoracic CT scans. We devise a similarity score and
show that it can be utilized to measure similarity 1) between annotations of the same
nodule by different radiologists and 2) between the query nodule and the top four CBIR
results. A comparison between doctors and algorithm scores suggests that the benefit
provided by the system to the radiologist end-user is comparable to obtaining a second
radiologist's opinion.

\keywords{CBIR \and Pulmonary Nodules \and Deep Learning \and Image Retrieval}
\end{abstract}
 \section{Introduction}
\subsection{Motivation and Background}

Lung cancer is a leading cause of cancer mortality in both men and women,
accounting for nearly 25\% of all cancer deaths\cite{Siegel2020}. The chances
of treating lung cancer successfully are much higher if the treatment starts at
an early stage.

Currently, Low-Dose Computed Tomography (LDCT) screening is the most effective
way for pulmonary nodules detection and diagnosis, and its usage has increased
dramatically over the last two decades. However, scan examination and diagnosis
is a very time-consuming task that requires a lot of invaluable radiologist
time.

To assist radiologists to quickly and effectively diagnose tumors, it is
important to present them with similar historical cases. Examination of similar
cases can be beneficial in two aspects. First, the radiologist can have access
to the labeling information that other doctors gave in similar cases and thus
can deduce the status of the current case. Second, the radiologist can examine
the related case development history past the similarly looking LDCT scan, as
if peeking at a possible future of the current case and infer a more accurate
prognosis. For example, a similar case biopsy outcome can suggest whether it is
advisable to perform a biopsy in the current case.

Many contemporary works that apply machine learning techniques to medical
imaging aim to replace the doctors and directly produce a diagnosis. The usual
downsides of this approach are the lack of the output interpretability and
uncertain robustness guarantees in the light of potential bugs, input
variation, or even adversarial attacks. Moreover, they are usually lacking in
the amount of diagnostic data they can provide, often limited by a handful of
bits of information, like benign-vs-malignant binary classification.

The solution we describe is a content-based image retrieval (CBIR) system.
Given a pulmonary nodule, our system retrieves several similar nodules from the
historical database, potentially enriched with the relevant clinical records,
to aid the doctor in the diagnostic process. Our main contributions include:

\begin{itemize}
\item We develop a system to provide the radiologist semantically-meaningful
decision support in nodule analysis, in contrast to providing an automated
nodule diagnosis.
\item We identify architectural constraints to our deep learning system and
provide theoretical justification for utilizing transfer learning.
\item We define a proxy task to learn the transformation of a 3D nodule to a
latent vector based on semantic features defined by medical experts.
\item We study semantic information preservation by our system, devise an
evaluation technique that considers the lack of consensus between the doctors,
and show that our method retrieves highly relevant results.
\end{itemize}

\subsection{Related Work}

Most of the previous research uses classic computer vision methods for feature
extraction, using a 2D representation of the nodules. This approach fails to
capture the full spatial nodule information. A significant challenge in CBIR
is the definition of the distance between two entities so that they will be
considered semantically similar in addition to being visually similar. Often,
the evaluation of the models is hindered by the lack of consensus between human
annotators. 

A few works in the field include the following: Lam et al.\cite{Lam2007}
performed CBIR on 2D slices of the 3D nodules using classical image
descriptors. Dhara et al.\cite{Dhara2016} extended this approach with manually
defined volumetric features. Pan et al.\cite{Pan2016} used spectral clustering
to transfer a 3D nodule to hash code used to retrieve similar nodules. Wei et
al.\cite{Wei2017} proposed a learned distance metric. Finally, Loyman \&
Greenspan\cite{Loyman2019} study included LIDC-IDRI rating regression from 2D
slices to obtain embeddings using deep learning.

 \section{Methods}
\subsection{Data}
\subsubsection{LIDC-IDRI.}

The Lung Image Database Consortium and Image Database Resource Initiative (LIDC-IDRI)
image collection consists of diagnostic and lung cancer screening thoracic computed
tomography (CT) scans with marked-up annotated lesions\cite{lidc_data,lidc_paper}. A panel
of four experienced radiologists performed independent segmentation and initial
categorization. Lesions categorized as nodules larger than 3mm in diameter were further
assessed for nine subjective characteristics: subtlety, internal structure, calcification,
sphericity, margin, lobulation, spiculation, radiographic solidity, and
malignancy\cite{McNittGray2007}. Each characteristic consists of either a discrete
category set or an integer rating on a five-point scale. We use pylidc
software \cite{Hancock2016} to access radiologist annotations.

Analysis of nodule characteristic distribution revealed four characteristics with very
low variability between nodules, which we decided to omit from further processing.
Adopting the LUNA16 \cite{Setio2017} approach we also limited the analysis to nodules accepted
by at least three out of four radiologists. This resulted in 1186 nodules, each with three
or four sets of segmentation and five-dimensional rating characteristic (subtlety,
sphericity, margin, lobulation, and malignancy). We normalize all rating values to $[0,1]$
range to aid the implementation.

\subsection{Methodology}

\subsubsection{CBIR Using Embeddings.}

Our approach to CBIR is partially inspired by the natural language processing technique
called word2vec\cite{Mikolov2013}. We learn a function $f:V\to S$ from a high dimensional
space of CT voxels $V$ to a much lower dimensional space $S$, which we call ``a semantic
space". A desirable property of $S$ is to capture nodule similarity as perceived by
radiology experts, that is, for some distance function $d:S^2\to\mathbb R$ we expect two
vectors $s_1, s_2 \in S$ to be relatively close (have small $d\left(s_1, s_2\right)$) if
the corresponding nodules would be considered similar by the doctors and vice versa. We
call vectors in $S$ ``embeddings". Not necessarily interpretable per se, we would like an
embedding to incorporate both characteristical information as defined by radiologists as
well as some visual information about the nodule. A traditional approach to learn
embeddings is to define a proxy task, such that training a deep learning model to solve
this task would produce the embeddings as a byproduct.

\subsubsection{Theoretical Architecture Constraints.}

According to PAC theory (e.g. see \cite[ch.~2]{Mohri2018}) 
bound of the generalization error over hypothesis space $\mathcal H$ and number of
training samples $N$ is:
\begin{equation}
R(h) \le \hat R_S(h) + O\left(\sqrt{\frac{\log\left|\mathcal H\right|}{N}}\right)
\end{equation}
Our dataset has a very limited number of nodules, while the deep learning network capable
of extracting the information from the 3D space of voxels can have many millions of
parameters, that is, $N\ll\log{|\mathcal H|}$. While some researchers question the
tightness of such theoretical generalization error bounds, we still wary of the model
overfitting in our setting. Therefore, we decide that our proxy task may not be sufficient
to train a robust feature extraction.

\subsubsection{Feature Extraction.}

Fortunately, thanks to LUNA16\cite{Setio2017} and Kaggle Data Science Bowl (DSB)
2017\cite{dsb2017} competitions there are a lot of previous works tackling pulmonary
nodule analysis. In this work we apply a transfer learning technique to the winning DSB
solution\cite{Liao2019} which is based on the U-net architecture\cite{Ronneberger2015}. It
has initially been trained on a dataset including lots of scans not found in LIDC-IDRI and
optimized for different tasks, namely, pulmonary nodule detection and binary whole-scan
classification. For our feature extraction, we reuse the pre-trained U-net backbone of the
nodule detector and drop the region proposal head comprising of the last two convolutional
layers.

\subsubsection{Rating Regression.}

\begin{figure}
    \centering
    \includegraphics[height=0.3\textheight]{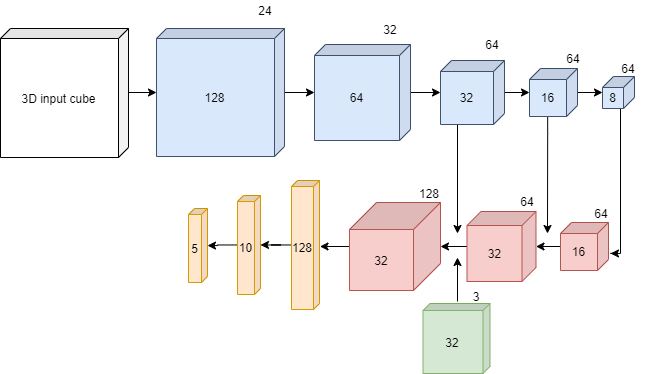}
    \caption{
        Regression network for embeddings learning (only feature map is shown). The
        trainable head comprises of the last three fully connected layers. See
        \cite{Liao2019} for a detailed explanation about the backbone.
    }
    \label{fig:regression}
\end{figure}

Here we define a proxy task as a radiologist rating regression. We extend the feature
extraction network with three fully connected layers with the output being a
five-dimensional vector (\autoref{fig:regression}). A target is defined as a mean of
radiologist ratings applied to the nodule at hand. We use MSE loss, training only the head
while keeping the backbone weights fixed. A ten-dimensional vector from the second-to-last
layer is used as an embedding.

 \section{Results and Discussion}
\subsection{Semantic Information Preservation}

In this section we study whether the embeddings produced by our method preserve
semantic meaning, that is, nodules with similar characteristics produce similar
embeddings.

\subsubsection{t-SNE.}

We run t-SNE\cite{Maaten2008} dimensionality reduction over the random sample
of the embeddings space (\autoref{fig:tsne}). For simplicity, the coloring here
only reflects malignancy: samples with high malignancy (mean rating $>3$ out of
5) are colored red and others blue. The separation of red and blue is not ideal
because embedding vectors preserve more information than just a malignancy
level. However, we observe that the malignant nodules are clustered together,
unlike the benign nodules. The more detailed analysis of malignant embeddings
reveals that the distance from the center of mass of the red cluster roughly
corresponds to the inverse malignancy rating of the nodule (not shown).

\subsubsection{Hierarchical Clustering.}

\begin{figure}
    \centering
    \begin{subfigure}[t]{0.49\textwidth}
        \includegraphics[width=\textwidth]{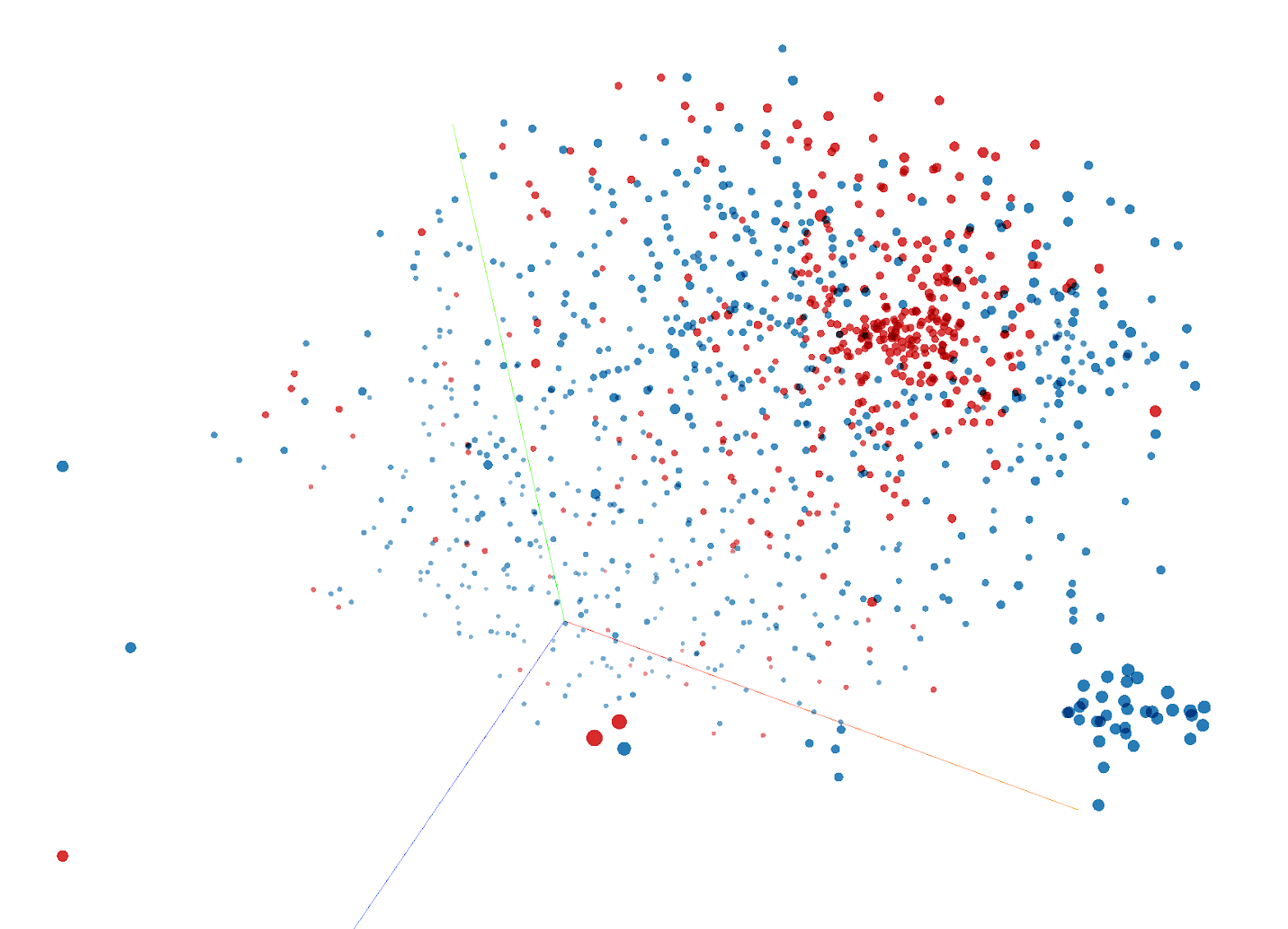}
        \caption{
            t-SNE visualization of embedding space. Embeddings of the malignant
            nodules (red) are clustered together, in contrast to the benign
            nodules (blue)
        }
        \label{fig:tsne}
    \end{subfigure}
    \hfill
    \begin{subfigure}[t]{0.49\textwidth}
        \includegraphics[width=\textwidth]{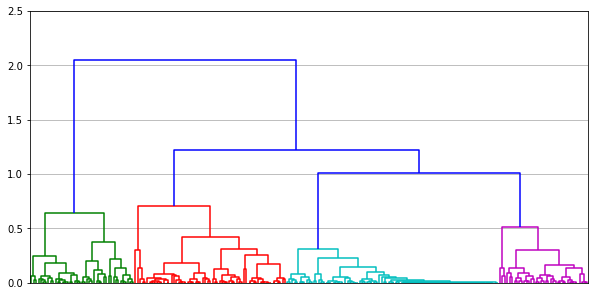}
        \caption{
            Hierarchical clustering of embeddings with top three splits
            corresponding to malignancy (green), margin (red) and subtlety
            (cyan, magenta) ratings
        }
    \label{fig:ward}
    \end{subfigure}
    \caption{Semantic information preservation in embedding space}
    \label{fig:seminfo}
\end{figure}

We also run a hierarchical clustering of a random subset of embeddings using
Ward's minimum variance method\cite{Ward1963} (\autoref{fig:ward}). We analyze
the top three splits and observe that the mean malignancy rating of nodules in
the leftmost group (green) is much higher than in other groups, that is, the
first split happens to cluster malignant vs benign nodules. Similarly, the
second split isolates low-margin benign nodules (red) from other benign ones,
and finally, the third split differentiates based on subtlety rating (among
benign nodules with high margin score).

\subsection{Usability: Top-$k$ Evaluation}

The purpose of our system is to aid a radiologist by presenting top-$k$ cases
similar to the query nodule. A radiologist can then inspect each presented
case, assessing morphological similarity and reviewing historical diagnosis or
even broader clinical records of the patients with presumably similar cases.
Aiming for interface simplicity and maximum user productivity we prefer a
minimum $k$ which still provides enough information to support a doctor's
decision. We study the semantic space distance between the retrieved samples
and a query and decide that beyond $k=4$ CBIR results provide diminishing
returns.

\subsection{Qualitative Assessment: Example Output}\label{sec:examples}

\begin{figure}
    \centering
    \begin{subfigure}[t]{\textwidth}
        \centering
        \includegraphics[height=3.9cm]{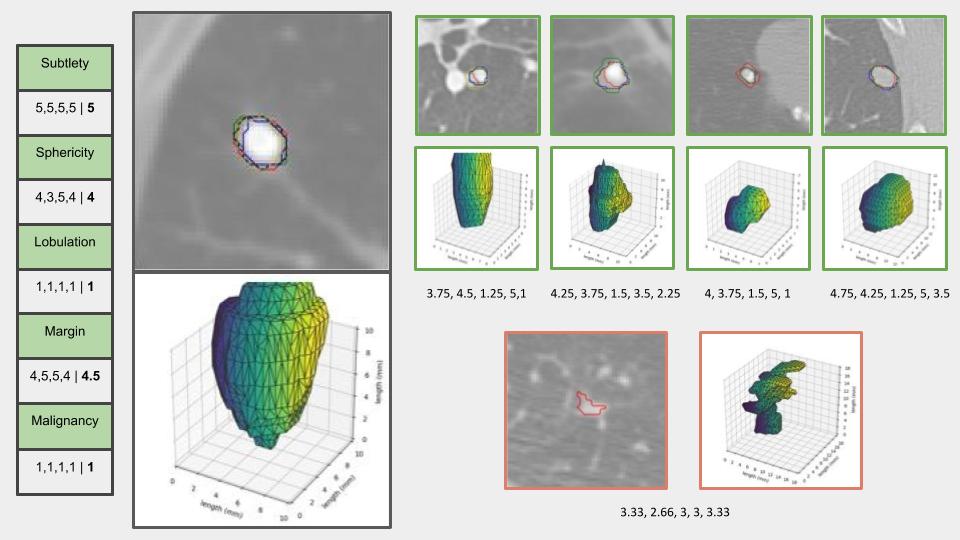}
        \caption{Low malignancy query example}
        \label{fig:ex1}
    \end{subfigure}
    \begin{subfigure}[t]{0.49\textwidth}
        \includegraphics[width=\textwidth]{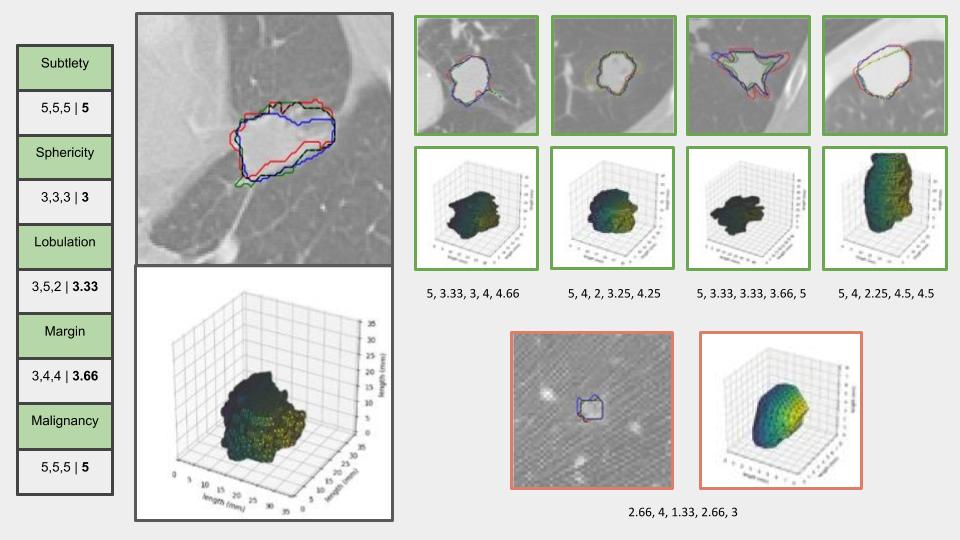}
        \caption{High malignancy query example}
        \label{fig:ex2}
    \end{subfigure}
    \hfill
    \begin{subfigure}[t]{0.49\textwidth}
        \includegraphics[width=\textwidth]{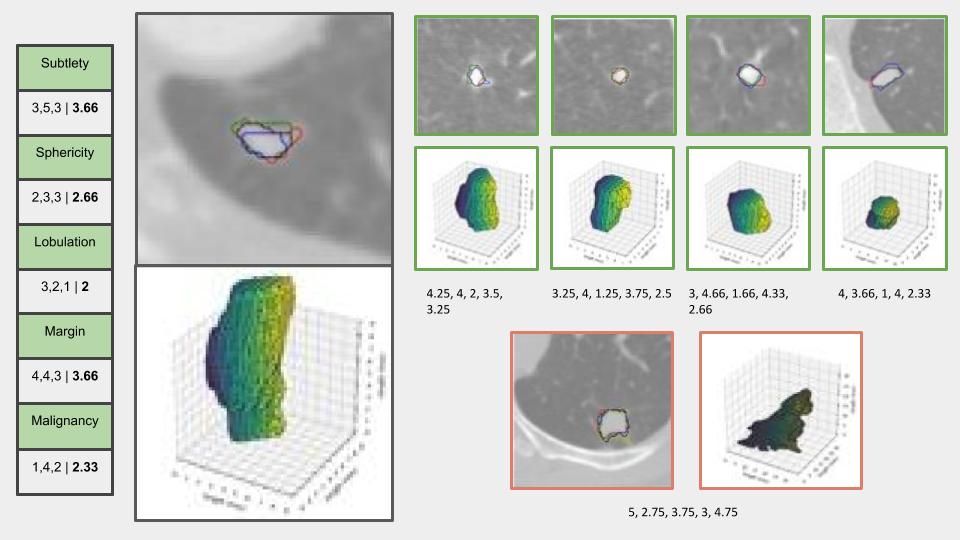}
        \caption{Uncertain malignancy query example}
        \label{fig:ex3}
    \end{subfigure}
    \caption{CBIR evaluation examples. See \autoref{sec:examples} text for the description.}
    \label{fig:examples}
\end{figure}

\autoref{fig:examples} shows example results of our system. The query nodule is
presented on the left side (gray border). Only a single CT slice is shown, with
doctors segmentation marks superimposed. A 3D mesh of consensus segmentation is
presented below the slice. Query ratings (before normalization) together with
the mean are provided on the left. The Top-4 CBIR results are shown on the
right (green border) together with their mean ratings. Notice the shape and
rating similarity of the CBIR results to that of the query. A non-match nodule
is displayed (red border) for comparison. We remind the reader that in contrast
to \autoref{fig:examples} visualization the CBIR query consists of a 3D CT
patch only.

\subsection{Quantitative Assessment}

To quantify the CBIR performance we conduct two evaluations: First, we compare
the CBIR-based algorithm to medical experts. Then we compare to a recent work
in the field \cite{Loyman2019}.

\subsubsection{Comparison to Human Experts.}

We define a nodule rating consensus as a mean vector of ratings assigned to the
same nodule by several radiologists. A \emph{dissent score} of the specific
radiologist ratings is defined as an RMSE between it and the consensus among
other radiologists. Similarly, a \emph{dissent score} of an algorithm is the
RMSE between ratings predicted by the algorithm and the consensus of all
radiologists (using normalized rating range $[0,1]$).

We use a rating consensus of $k$ top CBIR results as a na\"ive algorithm
prediction. We also provide, for comparison, a dissent score of a \emph{random
algorithm} that predicts one of the existing ratings randomly regardless of the
input.

\begin{figure}
    \centering
    \includegraphics[width=\textwidth]{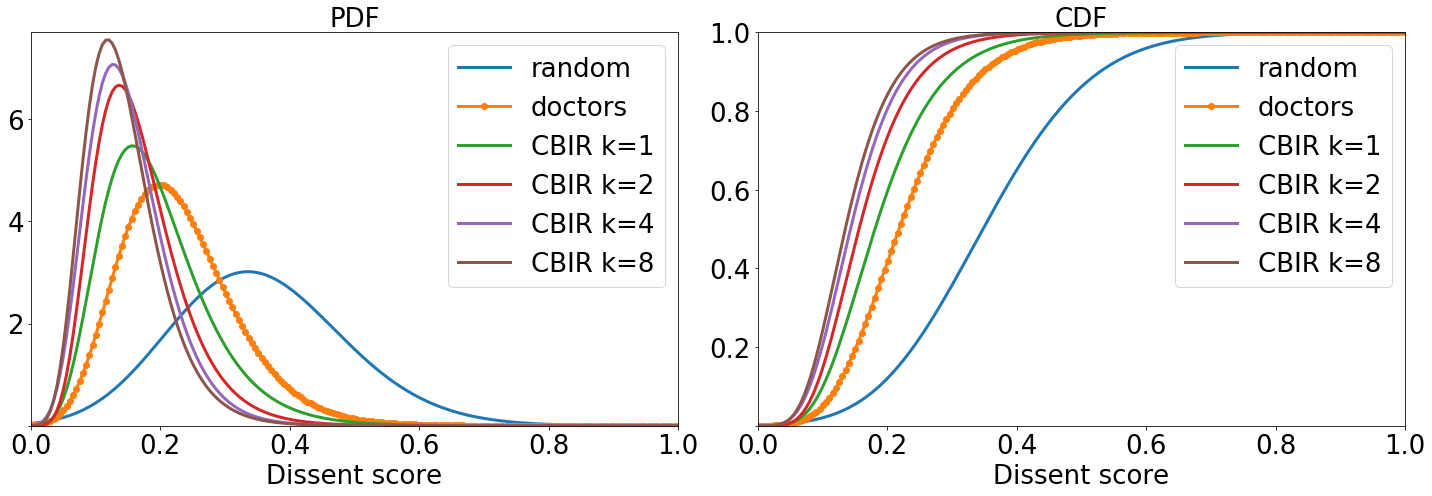}
    \caption{
        PDF and CDF of the dissent score of random predictions, doctor ratings,
        and na\"ive CBIR-based algorithm with $k\in \{1, 2, 4, 8\}$. For
        clarity, all graphs show MLE fit of the log-normal distribution. The
        lower dissent score is better.
    }
    \label{fig:alg_vs_doc}
\end{figure}

\autoref{fig:alg_vs_doc} shows a distribution of dissent scores, while
\autoref{tab:perf} reports their mean and standard deviation (measured over
five-fold cross-validation). We can see that the na\"ive CBIR-based algorithm
for $k=4$ improves on the doctors' diagnosis on average as much as the latter
improves on a random guess (1.5 times lower mean dissent score). That is, a
distance between the query and CBIR results is at least comparable to the
ranking uncertainty of the query itself, which is a strong indicator of the
embedding space quality.

\begin{table}
  \centering
  \begin{tabular}{l|cc|ccccc|ccccc|c}
  {} & \multicolumn{2}{c}{\parbox{4em}{\centering Dissent score}} & \multicolumn{5}{c}{Rating RMSE} & \multicolumn{5}{c}{Rating STD} & \\
  Method &    \rotatebox{90}{\scriptsize mean} &  \rotatebox{90}{\scriptsize STD} & \rotatebox{90}{\scriptsize subtlety} & \rotatebox{90}{\scriptsize sphericity} & \rotatebox{90}{\scriptsize lobulation} & \rotatebox{90}{\scriptsize margin} & \rotatebox{90}{\scriptsize malignancy} & \rotatebox{90}{\scriptsize subtlety} & \rotatebox{90}{\scriptsize sphericity} & \rotatebox{90}{\scriptsize lobulation} & \rotatebox{90}{\scriptsize margin} & \rotatebox{90}{\scriptsize malignancy} & \rotatebox{90}{Precision} \\
  \hline
  random     &    0.35 & 0.13 &     1.43 &       1.13 &       1.27 &   1.44 &       1.47 &     2.81 &       1.59 &       2.64 &   3.08 &       2.86 &           \\
  doctors    &    0.23 & 0.09 &     0.72 &       0.69 &       0.74 &   0.73 &       0.75 &     0.83 &       0.64 &       0.98 &   0.95 &       0.85 &           \\
  CBIR k=1   &    0.19 & 0.08 &     0.89 &       0.79 &       0.84 &   0.94 &       0.87 &     0.59 &       0.48 &       0.58 &   0.61 &       0.56 & \bf     0.81 \\
  CBIR k=2   &    0.17 & 0.07 &     0.76 &       0.70 &       0.74 &   0.81 &       0.73 &     0.49 &       0.41 &       0.48 &   0.52 &       0.47 &      0.80 \\
  CBIR k=4   &    0.15 & 0.06 &     0.70 &       0.65 &       0.68 &   0.74 &       0.68 &     0.45 &       0.38 &       0.44 &   0.46 &       0.44 &      0.80 \\
  CBIR k=8   & \bf   0.14 & \bf0.06 & \bf    0.66 & \bf      0.63 & \bf      0.64 & \bf  0.70 & \bf      0.64 & \bf    0.43 & \bf      0.37 &       0.42 &   0.44 & \bf      0.41 &      0.79 \\
  CBIR mean &         &      &          &            &            &        &            &          &            &            &        &            &  \bf 0.79 \\
  Loyman et al. &         &      &     0.93 &       0.83 &       0.89 &   0.94 &       0.68 &     0.84 &       0.47 & \bf      0.27 & \bf  0.37 &       0.84 &      0.75 \\
  \end{tabular}
  \caption{
       Quantitative comparison of methods: Random guess, human expert,
       presented CBIR solution (``CBIR"), and Loyman et al \cite{Loyman2019}.
       For the RMSE-based metrics, as defined here and in \cite{Loyman2019},
       lower is better. For Retrieval precision, higher is better. The best
       results are in bold.
  }
  \label{tab:perf}
\end{table}

\subsubsection{Comparison to Recent Results.}

We extend \autoref{tab:perf} with a comparison to \cite{Loyman2019}. Following
the methodology defined in \cite{Loyman2019}, we compute a prediction RMSE and
STD over the dataset for each rating component separately (using rating range
$[1,5]$) and compare to the rating regression results from \cite{Loyman2019}.
We also compute a CBIR malignancy retrieval precision, that is: what portion of
the retrieved $k$ results has a correct binary malignancy class (benign vs
malignant). Since \cite{Loyman2019} only presents mean precision for $k\in\{1,
3, 5, 7, 9, 11, 13, 15\}$ we compute it as well for a meaningful comparison
("CBIR mean" in the Table). From the results shown in the Table, we can see
that our CBIR-based rating prediction improves over \cite{Loyman2019} by 0.20
for RMSE and by 0.14 for STD on average. It also compares favorably to human
experts. The retrieval precision is improved by 4\%.

 \section{Conclusion}

In this work we prototype a CBIR system for pulmonary nodules. We develop a methodology to
learn low-dimensional embeddings and present a theoretical justification for the
architectural design selected. Our methodology facilitates learning of high-quality
embeddings preserving both spatial and semantic information, all this despite very high
data dimensionality and sample scarcity. We determine optimal usability settings and
perform a qualitative analysis of CBIR results.  Finally, we conduct a quantitative study
demonstrating state-of-the-art results. While we believe that in reality, the diagnostic
process is more complex than a 5-tuple rating can convey, we conclude that the CBIR output
is highly relevant to the query, such that the benefit provided by the system to the
radiologist end-user is comparable to obtaining a second radiologist's opinion.

\subsubsection{Acknowledgments.}
Part of the work presented in this paper was done by the first two authors in
the course of Y-Data program by Yandex School of Data Analysis. The authors
would like to thank Kostya Kilimnik and Shlomo Kashani for the organization and
support of the project initiation.

\bibliographystyle{splncs04}

\end{document}